\documentclass[usenatbib,useAMS,a4]{mn2e}
\usepackage[dv ips]{graphicx}
\usepackage{natbib}
\usepackage{lastpage}
\usepackage{subfigure} 

\title{The evolution of planetesimal swarms in self-gravitating protoplanetary discs}
\author[Joe Walmswell, Cathie Clarke and Peter Cossins]{Joe Walmswell$^1$ \thanks{E-mail: jjw49@ast.cam.ac.uk}, Cathie Clarke$^1$ and Peter Cossins$^2$\\
$^1$Institute of Astronomy, Madingley Rd, Cambridge, CB3 0HA, UK\\
$^2$Dept. of Physics and Astronomy, University of Leicester, Leicester, LE1 7RH}

\date{February 2013}
\pagerange{\pageref{firstpage}--\pageref{lastpage}}
\begin{document}

\maketitle
\label{firstpage}

\begin{abstract}
We investigate the kinematic evolution of planetesimals in
self-gravitating discs, combining Smoothed Particle Hydrodynamical
(SPH) simulations of the disc gas with a gravitationally coupled
population of test particle planetesimals. We find that at radii of
10s of au (which is where we expect planetesimals to be possibly
formed in such discs) the planetesimals' eccentricities are rapidly
pumped to values $>$ 0.1 within the timescales for which the disc is
in the self-gravitating regime. The high resulting velocity dispersion
and the lack of planetesimal concentration in the spiral arms means
that the collision timescale is very long and that the effect of those
collisions that do occur is destructive rather than leading to further
planetesimal growth. We also use the SPH simulations to calibrate
Monte Carlo dynamical experiments: these can be used to evolve the
system over long timescales and can be compared with analytical
solutions of the diffusion equation in particle angular momentum
space. We find that if planetesimals are only formed in a belt at
large radius then there is significant scattering of objects to small
radii; nevertheless the majority of planetesimals remain at large
radii. If planetesimals indeed form at early evolutionary stages, when
the disc is strongly self-gravitating, then the results of this study
constrain their spatial and kinematic distribution at the end of the
self-gravitating phase.
\end{abstract}


\section{Introduction}

There are commonly discussed mechanisms by which gas giant planets can form
out of the gas and dust of the protoplanetary disc. On the one hand,
part of the disc may become sufficiently dense to become
gravitationally unstable and
collapse \citep{2000ApJ...536L.101B}. Alternatively, in what has become
to be known as the `core accretion' scenario, the dust in the disc
coagulates to form ever larger objects, the planetesimals, eventually
resulting in a solid core that is large enough to initiate runaway
accretion of a gaseous envelope (e.g. \citet{1996Icar..124...62P},
\citet{2005Icar..179..415H}). In the latter scenario planetesimal coagulation must occur while the disc is still gas rich
(in contrast to the case of the assembly of terrestrial planets, where
the process may occur after dispersal of the disc
gas). Unsurprisingly, therefore, there is a large literature devoted
to planetesimal evolution in the presence of gas
(\citet{1976PThPh..56.1756A},\citet{1996Icar..124...62P},
\citet{2004ApJ...602..388T}). In most cases, the gas is treated as a
laminar flow although several authors (\citet{2004ApJ...612L..73L},
\citet{2004MNRAS.350..849N}, \citet{2005A&A...443.1067N},
\citet{2008ApJ...686.1292I}, \citet{2010MNRAS.409..639N}) have considered instead the scenario
where planetesimals are subject to stochastic torques arising from
fluctuations in a disc subject to turbulence generated by the magneto-rotational instability (MRI).  In the case of laminar discs,
planetesimal eccentricity is (weakly) pumped by mutual gravitational
scattering (e.g. \citet{1969QB981.S26......}) and (weakly) damped by
gas drag, so that the equilibrium distribution of eccentricities is
peaked at low values ($\sim 0.01$). This low value is important to
continued planetesimal growth since it implies low relative velocities
for planetesimal encounters: this not only favours agglomerative (as
opposed to destructive) outcomes (\citet{1999Icar..142....5B},
\citet{2008ssbn.book..195L}) but also enhances the collision cross-section ($\propto e^{-2}$) in the gravitationally-focused regime.

  All of the above studies treat the evolution of planetesimals in
  {\it non}-self-gravitating discs. However,
  \citet{2004MNRAS.355..543R} have argued that planetesimals may be
  formed very early (in the first few $10^5$ years) of a disc's life
  when it is still strongly self-gravitating; at this stage, spiral
  features in the disc provide pressure maxima in which dust is
  concentrated through the action of gas drag. The enhanced collision
  rates are favourable to grain growth and \citet{2006MNRAS.372L...9R}
  have argued that self-gravity in the {\it dust} phase may even
  promote the formation of km scale structures
  (i.e. planetesimals). Observational evidence for at least the early
  stages of grain growth during the self-gravitating phase is provided
  by the detection of $10$ cm radiation from HL Tau
  \citep{2008MNRAS.391L..74G}, which implies that the growth of grains
  to at least cm scales (from the sub-micron scales typifying the
  interstellar medium) has already occurred in this young and massive
  disc system.

 If planetesimal formation indeed belongs to the earliest (self-
 gravitating) phase of disc evolution, then it is necessary also to
 trace the evolution of planetesimals during the self-gravitating
 phase.  The dynamical evolution of planetesimals in this environment
 has a number of implications for planet formation and for the
 collisional production of dust in disc systems. For example, the
 relative importance of collisional growth of planetesimals in the
 self-gravitating and non-self gravitating phase of the disc can be
 crudely assessed (cf \citet{2008MNRAS.385.1067B}) by comparing the
 product of the lifetime of each phase, the typical disc density and
 the (inverse) square of the typical planetesimal velocity
 dispersion. The first two terms roughly cancel (i.e discs typically
 live an order of magnitude longer in the non-self gravitating phase
 but with disc masses an order of magnitude lower) so the relative
 importance of collisional growth in the two regimes boils down to the
 relative values of the velocity dispersion. 

A pilot study of the dynamical evolution of planetesimals in
self-gravitating discs \citep{2008MNRAS.385.1067B} demonstrated that
high orbital eccentricities ($e$) are generated in such discs with
particles undergoing stochastic changes in their orbital elements as a
result of interaction with spiral features in the disc: high
eccentricities imply a high velocity dispersion (of order $e v_k$
where $v_K$ is the local Keplerian velocity) if the particle
trajectories are randomly phased. A lower velocity dispersion would
however apply if the planetesimals instead demonstrated local velocity
coherence: the small number of particles modeled by
\citet{2008MNRAS.385.1067B} did not permit exploration of this
possibility however. For the same reason, it was not possible to
discern the sign of any net migration of the planetesimal swarm nor
whether the main statistical effect involved changes in orbital energy
or angular momentum. All these issues have implications for
planetesimal growth during the self-gravitating phase and also for the
retention of planetesimals against loss through radial migration
\citep{2005ApJ...627..286T} or collisional grinding to small dust
\citep{2007ApJ...658..569W}. Moreover, evolution during the
self-gravitating phase controls the spatial distribution and dynamical
properties of any planetesimals that survive this phase.

 All of the above provides ample motivation for the present study in
 which we explore the response of a large ensemble of planetesimals to
 the gravitational torques imposed by a self-gravitating gas disc. We
 restrict ourselves to the regime where planetesimals respond as test
 particles to the imposed potential fluctuations. This in practice
 restricts us to sizes $>$ km scale (in order to be able to
 neglect gas drag; \citet{2008MNRAS.385.1067B}) and $< 1000 $ km scale
 (in order to be in the test particle regime where the effect of
 gravitational perturbations induced by the particles in the disc gas
 can be ignored; \citet{2003MNRAS.341..213B}).

In Setion 2 we present a numerical investigation (modeling the disc
with Smoothed Particle Hydrodynamics (SPH)) that extends the pilot
study of \citet{2008MNRAS.385.1067B} to a large ensemble of
planetesimals. Our results (Section 3) imply the disc pumps particle
eccentricity with only modest changes in the semi-major axis and this
motivates the analytical modeling and Monte Carlo simulations that we
present in Section 4. This analytical/Monte Carlo approach is able to
reproduce the results of the SPH simulation over the rather limited
time frame that is possible in the latter case and also permits
integration over long timescales. In Section 5 we discuss the results
of these calculations in relation to the questions posed above and
summarise our conclusions.

\section{The simulation}

\subsection{The physics of self-gravitating discs}

 We model the gas disc as a self-gravitating disc which achieves a
 situation of {\it self-regulation}: i.e. it settles to a
 marginally-stable state where the Toomre $Q$ parameter
 \citep{1964ApJ...139.1217T} obeys

\begin{equation}
\ Q = \frac{c_{{\rm s}}\kappa}{\pi G \Sigma} \sim  1.
\end{equation}
 
 Here $c_s$ is the local sound speed, $\Sigma$ the gas surface density
 and $\kappa$ is the epicyclic frequency, which is $\sim \Omega$ (the
 Keplerian frequency) for the low mass discs considered here.  In this
 state the disc is in a state of thermal equilibrium between the
 heating associated with the gravitational instability and the imposed
 cooling.  Here we follow \citet{2001ApJ...553..174G} and
 \citet{2005MNRAS.358.1489L} in parameterising the cooling in terms of
 a cooling time that is a fixed multiple ($\beta$) of the local
 dynamical time ($\Omega^{-1}$), i.e.  we have ${t_{{\rm cool}} = \beta
   \Omega^{-1}}$ and

\begin{equation}
Q_{-} = -\frac{u}{t_{{\rm cool}}} \label{eq:cooling}
\end{equation}
 
where ${Q_{-}}$ and $u$ are the cooling rate per unit mass and
thermal energy per unit mass.

The parameter $\beta$ is a measure of overall cooling and the choice
of $\beta$ determines the magnitude of the spiral density waves
required to maintain thermal equilibritum. By equating this cooling
law with the predicted heating rate per unit mass ${Q_{+}}$ from
the instability, \citet{2009MNRAS.393.1157C} demonstrated that the
fractional density perturbation, a measure of relative spiral arm
strength, is proportional to $\beta^{-1/2}$; the numerical simulations
in the same study confirmed this dependency and suggested that

\begin{equation}
\frac{\delta\Sigma}{\Sigma} \approx \frac{1}{\sqrt{\beta}} \label{eq:ampl}.
\end{equation}

This means that in our constant $\beta$ simulations we would expect
the density perturbation to be approximately constant across the disc
and to increase as $\beta$ decreases. In practice, if $\beta$ is less
than about 4 in a simulation, the magnitude of the perturbation
becomes non-linear and the disc fragments
\citep{2001ApJ...553..174G}. (See e.g. \citet{2011MNRAS.410..559M},
\citet{2012arXiv1209.1107M}, \citet{2011MNRAS.413.2735L},
\citet{2011MNRAS.416L..65P}, \citet{2012MNRAS.421.3286P} and \citet{2012MNRAS.420.1640R}) for an
ongoing debate as to whether fragmentation may also eventually occur
in well resolved simulations at significantly higher values of
$\beta$.)

\subsection{Simulation parameters}

 We model the system as a point mass, mass $M$, orbited by 250,000 SPH
 gas particles and 50,000 test particles; for details of the code see
 \citet{2009MNRAS.393.1157C} and references therein. The test
 particles are assigned masses equal to $10^{-6}$ times the mass of a
 gas particle and are subject only to gravitational forces.  Gas
 particles are accreted onto the point mass if they enter within a
 sink radius of 0.25 code units and satisfy certain conditions
 \citep{1995MNRAS.277..362B}; the point mass itself is free to
 move. By the end of the simulation, no more than 2 per cent of the
 gas particles are accreted.

Artificial viscosity is included, according to the standard SPH
formalism, with $\alpha_{{\rm SPH}}=0.1$ and $\beta_{{\rm SPH}}=0.2$; for the
parameters employed, the ratio of SPH smoothing length to disc
vertical scale height is about two throughout. The gas is
modelled as a perfect gas with $\gamma =5/3$ and is subject to both
compressive heating and shock dissipation.  The cooling law
(Equation~\ref{eq:cooling}) means that heat lossq depends only on the
dimensionless parameter $\beta$, so cooling is scale free.  For the
main simulation we adopt $\beta = 5$ which corresponds to rather large
amplitude spiral perturbations (see Equation~\ref{eq:ampl}); we contrast these
results with the lower amplitude case where $\beta =10$.
  
We employ scaled units so that $G=M=1$ and express time in units of
the dynamical time ${t_{{\rm dyn}}=\Omega^{-1}}$ at $R$=1. This time
$t$ always refers to the time since the start of the simulation.
We use these dimensionless results because they can be easily scaled to real
situations. Our standard model consists of a disc with total mass
equal to $0.1$, distributed with surface density $\propto R^{-3/2}$
over the radial range $1$ to $25$ and the planetesimals are
distributed according to the same radial distribution; we also
investigate a similar set-up but with $\beta =10$ and total mass of
$0.5$. We initialise the simulations with uniform temperature so that
$ Q=2$ at all radii in the disc. At each radius we distribute the
particles in $z$ in a Gaussian distribution representing approximate
hydrostatic equilibrium at the initial temperature. True hydrostatic
equilibrium is reached within a few local dynamical times.

\section{Simulation Results}

 The disc is initialised with ${Q=2}$ throughout and is thus
 initially gravitationally stable. It then cools on the local cooling
 time until it attains ${Q~1}$, initiating the gravitational
 instability and liberating heat. By time $t=4000$ the disc has definitely settled
 into a quasis-steady state where spiral features form and dissolve on a
 roughly dynamical timescale and with $Q~1$ for $5<R<25$. (At $R <
 5$ the value of $Q$ rises above unity since the disc is poorly
 resolved and heating by artifical viscosity acting on the Keplerian
 velocity field is sufficient to maintain the disc in a
 gravitationally stable state.)  Since this is a purely numerical
 artefact, we restrict our attention to the regime $R > 5$.

\subsection{Escapees}

The first consideration is whether the test particles of the swarm
gained enough energy from their interactions with the disc to escape
the system entirely.  We assess this by calculating the energies of
the test particles (for this purpose we approximate the potential due
to the disc by simply - for each particle - adding the enclosed disc
mass to the effective mass of the central object and ignoring the
contribution from external particles; although this is not strictly
correct in a disc system, this does not significantly impair our
ability to differentiate between bound and unbound particles).

  We find that (once the disc has settled into a steady gravoturbulent
  state) around $200$ particles (out of a total of $10^5$) escape
  during the subsequent $4000$ time units. If we (somewhat
  aribitrarily) equate code units for mass and radius with $1 M_\odot$
  and $1$ au then this implies an e-folding timescale of $ 3 \times
  10^5$ years (i.e. of order the self-gravitating lifetime of the
  disc). This e-folding timescale is likely to be an under-estimate
  for two reasons: the low value of $\beta$ employed in the
  simulations (which implies much more vigorous spiral activity than
  would be expected given the relatively long cooling timescales in
  realistic protostellar discs; \citet{2009MNRAS.396.1066C} and
  \citet{2007A&A...475...37S}) and the assumption of constant disc
  mass (whereas in reality the disc mass would decline over such
  timescales). {\footnote {We found an approximately $3$ times higher
      escape rate in the case of our massive disc simulation (with
      disc mass five times higher than the standard case), which
      demonstrates that - as expected - the rate of test particle
      escape would drop as the disc mass declines.}}  On this basis,
  we conclude that the majority of planetesimals would in reality
  remain bound.

\subsection{Ring spreading}

\begin{figure}
\centering
\includegraphics[width=0.99\columnwidth]{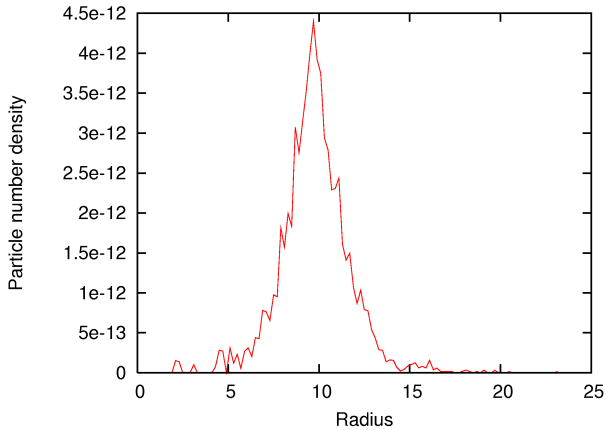}
\caption{Ring profile for $R=10$ after five local periods and with initial width of 1 at time $t=4000$.}
\label{Ring10}
\end{figure}

\begin{figure}
\centering
\includegraphics[width=0.99\columnwidth]{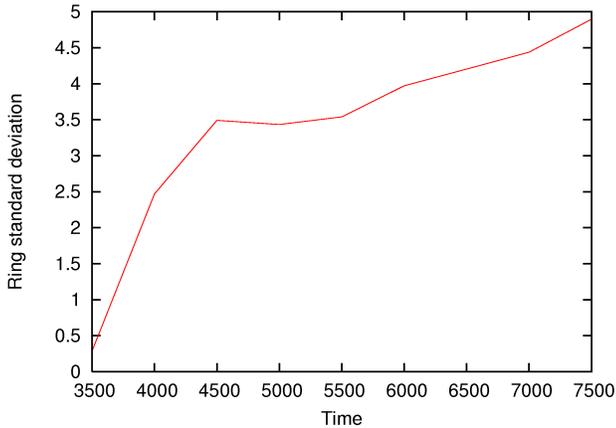}
\caption{Evolution of the ring standard deviation at $R=20$. The ring is selected at $t=3500$, when the disc is in the quasi-steady state, and then followed for the rest of the simulation.}
\label{Sdtime}
\end{figure}

Once the system has settled into the equilibrium state (i.e.  after $t= 3-4000$) we identify rings of particles over a restricted radial
range and analyse their spatial evolution. Figure \ref{Ring10}
illustrates the rather rapid spreading after five local orbital
periods of a ring of initial width $1$, initially located at $R=10$ at
$t=4000$. We find that the fractional change in the centroid of the
distribution and the change in the width of the distribution
(normalised to the initial radius) is independent of the radius of the
ring selected as long as the comparisons are conducted after the same
number of local dynamical timescales.  This is as expected, given that
(for the constant $\beta$ that we impose) the fractional amplitude of
the gravitational perturbations should be independent of radius. Note
that although the mode of the particle radii has moved in slightly, the
mean has moved out. Figure \ref{Sdtime} illustrates the evolution of
the fractional standard deviation of a ring at $R=20$ selected at
$t=3500$ and illustrates the continued spreading of the ring over many
local orbital times.

\begin{figure}
\centering
\includegraphics[width=0.99\columnwidth]{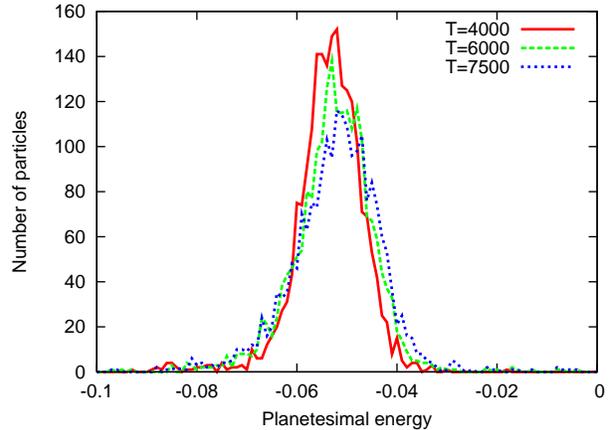}
\caption{Evolution of the energy distribution of the R=10 ring, selected at t=4000.}
\label{Energy}
\end{figure}

 We find that the rings spread in response to stochastic interaction
 with the disc and also that there is a small outward shift in the
 particle centroid for each ring. We can understand this behaviour by
 considering the evolution of the energy distribution of particles in
 a selected ring (Figure \ref{Energy}): the finite width of the
 initial distribution just reflects the finite width of the ring. It
 is notable that there is little evolution in the distribution of
 particle energies over the time period of $t=4000$ to $t=7500$,
 whereas ring spreading is still significant over this period (Figure
 \ref{Sdtime}). Although this spreading can be partially attributed to
 phase mixing (given the finite eccentricities of the particles
 selected), this is not the only process at work, as is illustrated by
 Figure \ref{EM}, which shows the secular increase of the particle
 eccentricities. This suggests that as an ensemble, the particles are
 undergoing little evolution in energy but are diffusing in angular
 momentum. This implies a growth of particle eccentricity which we
 quantify below for the entire disc.  (It should be stressed that the
 constancy of energy - and hence semi-major axis - is a property of
 the local ensemble; as was noted in \citet{2008MNRAS.385.1067B},
 individual particles undergo stochastic interactions in which they
 undergo modest changes in energy).

\subsection{Eccentricity growth}

\begin{figure}
\centering
\includegraphics[width=0.99\columnwidth]{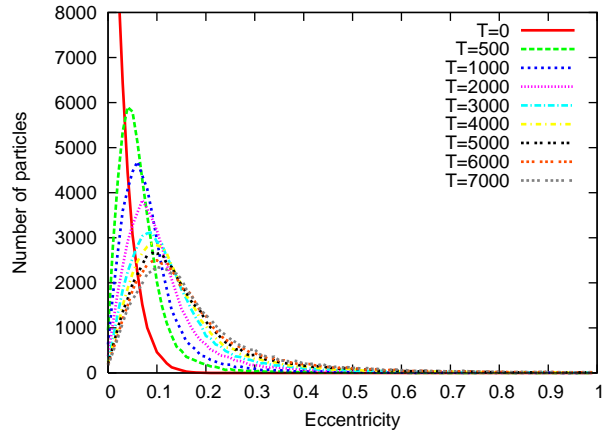}
\caption{Eccentricity evolution of the standard simulation.}
\label{E}
\end{figure}

 Figure \ref{E} depicts the eccentricity distribution for the entire
 disc ensemble over the duration of the simulation and confirms the
 evolution towards higher eccentricity, as would be expected if the
 spatial spreading largely reflects the growth of particle
 eccentricity at fixed semi-major axis. Note that in the period before
 the disc reaches the quasi-steady state (i.e. between $t=0$ and
 $t=3-4000$) the eccentricity distribution is not being pumped by the
 recurrent spiral features. There does not appear to be a sharp
 transition between the two regimes though. 

\begin{figure}
\centering
\includegraphics[width=0.99\columnwidth]{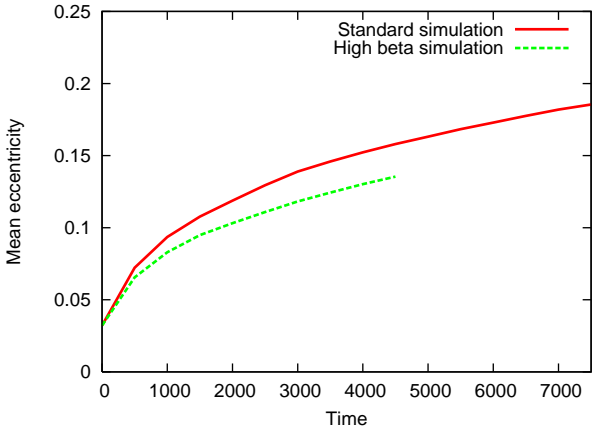}
\caption{Evolution of the eccentricity means for the two simulations.}
\label{EM}
\end{figure}

In Figure \ref{EM} we compare the evolution of the mean eccentricity
between the standard simulations and the high $\beta$ case (in which
$\beta$ differs by a factor of $2$). We see that the timescale to
attain a given mean eccentricity is roughly doubled in the higher
$\beta$ case. This is consistent with a picture in which the particles
are responding diffusively to gravitational interactions with the
disc: the diffusion timescale scales with the inverse of the diffusion
coefficient and thus with the inverse square of the amplitude of the
perturbations. Hence (from Equation~\ref{eq:ampl}) we expect the
timescale for eccentricity growth to scale linearly with $\beta$, as
is consistent with Figure \ref{EM}. We note that the mean eccentricity
increases smoothly after $t=500$ and that, as with the distribution as
a whole, we cannot identify a sharp transition when the quasi-steady
state is reached. This means that we can consider the distribution to
be a reasonable proxy for how it would appear had the disc been in the
quasi-steady state throughout, a fact we use in Section 4 to scale our
analytical model and Monte Carlo simulations.

\subsection{The velocity dispersion}
\begin{figure}
\centering
\includegraphics[width=0.99\columnwidth]{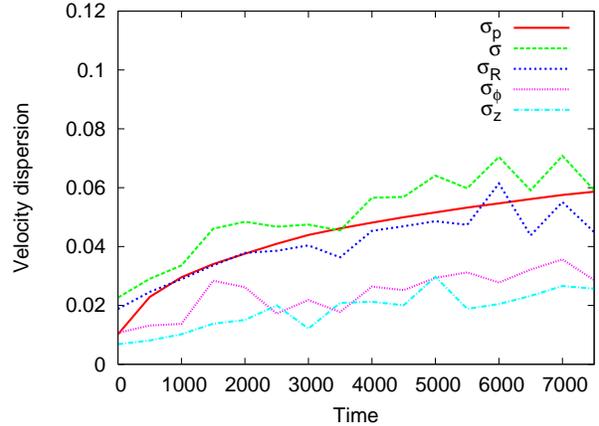}
\caption{Evolution of the velocity dispersion ($\sigma$) and its
  components in the radial ($\sigma_R$), azimuthal ($\sigma_{\phi}$),
  and vertical ($\sigma_z$) directions, compared with the prediction
  for randomly orientated orbits ($\sigma_p$).}
\label{V}
\end{figure}

For a swarm of particles with randomly orientated orbits, the velocity
dispersion is approximately equal to the product of the eccentricity
and the local Keplerian velocity \citep{2008MNRAS.385.1067B}. The
strong growth in particle eccentricity would thus raise the velocity
dispersion of the planetesimals with consequences for particle growth
that we discuss below. However, this would be an over-estimate of the
velocity dispersion if there was a degree of local velocity coherence
within the particle swarm. We investigate this in Figure~\ref{V} by
comparing the evolution of the local velocity component ($\sigma$) and
its three orthogonal components (computed within a patch of disc
containing $\sim 100$ particles) with the product of the instantaneous
mean eccentricity and the local Keplerian velocity ($\sigma_{{\rm p}}=e
v_{\rm k}$). The figure illustrates that $e v_{\rm k}$ is indeed a good measure of
the local velocity dispersion, as is expected in the case of randomly
phased elliptical orbits. In addition, the ratio between the radial
and azimuthal dispersions is maintained at about 3:2. One can show
from the epicyclic approximation for a cylindrical potential that the
expected ratio between the dispersions should be
$\sigma_{\phi}^{2}=0.5\sigma_{R}^{2}$ \citep{2008gady.book.....B},
i.e. that $\sigma_{\phi}=\sigma_{R}/\sqrt{2}$ or $3 \sigma_{\phi}
\approx 2 \sigma_{R}$.

 The growth of the velocity dispersion has profound
 consequences. First, a high velocity dispersion suppresses the role
 of gravitational focusing in facilitating collisions
 \citep{2008MNRAS.385.1067B}. In the limit that the disc scale height
 exceeds the mean separation between planetesimals the timescale for
 physical collisions between planetesimals of mass M, radius R and
 density $\rho$ is

\begin{equation}
t_{{\rm grow}} \sim \frac{\rho^{2/3} M^{1/3}}{\Sigma_p\Omega(1+4 G M / (R \sigma^{2}))},
\end{equation}

where $\Sigma_{\rm p}$ is the planetesimal surface density
and $\sigma$ is the planetesimal velocity dispersion. The $\sigma$
dependence means that collisions are highly disfavoured. At 30 au from
a solar mass star and using ${ e=0.2}$, the velocity dispersion is
around 1 km${\rm s^{-1}}$. Using the above formula for 100 km
planetesimals, the collision timescale is of order $10^{9}$ years! Smaller
planetesimals would give even longer timescales. Since planetesimal
growth cannot proceed without physical collisions, this rules out
significant planetesimal growth during the self-gravitating phase.

In addition - even when collisions do occur - the collisional velocity
is correspondingly high. The velocity required to shatter
planetesimals by collision is a few 100 m${\rm s^{-1}}$
\citep{2008ssbn.book..195L}, meaning that collisions would tend to result in
disruption rather than in growth. We can then use the above collision
probabilities to estimate a rough upper limit on the amount of small
dust (i.e. in the observable regime of $<1$mm) that could be generated
by disruptive collisions between planetesimals. On the optimistic
assumption that every planetesimal-planetesimal collision results in
all its mass being liberated as dust, the fraction of mass in
planetesimals that can be liberated is simply the ratio of the
self-gravitating lifetime to the collision time, i.e. $\sim
10^{-4}$. We therefore conclude that the role of planetesimal
collisions in re-supplying small dust is insignificant during the
self-gravitating phase. Once the disc enters the non-self gravitating
phase, and gas drag reduces the equilibrium eccentricity level to around
0.01 \citep{2000Icar..143...15K}, the collision rate rises by around a
factor of $100$ and at this stage either collisional growth and/or
dust production may become important.

\subsection{Planetesimal concentration in the arms}

 The above estimates for planetesimal collision frequencies neglect
 any possible concentration of particles in spiral features. For the
 values of $\beta$ employed in the simulations, the gas surface
 density varies by no more than a factor of two between the arms and
 the inter-arm regions (as expected, since we have modeled a regime
 where the disc is close to - but not at - the fragmentation
 boundary). If the planetesimals simply followed the gas, such an
 enhancement would have only a minor effect on planetesimal collision
 probabilities, even in the unlikely event that planetesimals spent
 all their time in long-lived regions of density enhancement. In fact
 we find that - instead of being preferentially concentrated in the
 arms - the planetesimal surface density varies by no more than $\sim
 20 \%$ around the orbit in the standard simulation, i.e. with
 amplitude much less than that of the gas.{\footnote {Note that the
     initial process of concentrating solid material into spiral
     features described by \citet{2004MNRAS.355..543R} - which is
     required for the formation of planetesimals in the
     self-gravitating phase - instead involves the effect of gas drag
     on small particles, and this, by definition, is ineffective in
     the test particle regime considered here.}}

 \citet{2008MNRAS.385.1067B} suggested that there were possible hints
 that the concentration of particles in the arms was more efficient in
 the case of more massive discs. In their simulation with a disc mass
 of half that of the point mass they identified epochs at which some
 test particles would settle into orbits that co-rotated with the
 spiral modes and conserved a Jacobi constant.  We have analysed our
 `massive disc' simulation and however find that, as in the standard
 simulation, there is no discernible influence on the planetesimal
 distribution: there is no significant concentration of planetesimals
 in the arms.

\subsection{A truncated planetesimal distribution}

  We finally consider the case that the planetesimal distribution is
  truncated, i.e. that it does not initially extend within a certain
  inner radius, $R_{{\rm in}}$, and study the extent to which planetesimals
  are scattered into regions within $R_{{\rm in}}$. This choice is motivated
  by the suggestion of \citet{2009MNRAS.398L...6C} that planetesimals
  are only likely to form in self-gravitating discs at large radius
  (beyond a few $10$s of au). We also note that the simulations of
  \citet{2012MNRAS.426.1444G} suggest that the mechanism is probably
  restricted to radii greater than 20 au. At such radii the cooling
  time is relatively short (corresponding to $\beta < 10$;
  \citet{2009MNRAS.396.1066C}) and the amplitude of spiral
  disturbances is large enough for rapid concentration of dust in
  spiral structures (i.e.  on timescales shorter than the (roughly
  dynamical) lifetimes of spiral features).
  
\begin{figure}
\centering
\includegraphics[width=0.99\columnwidth]{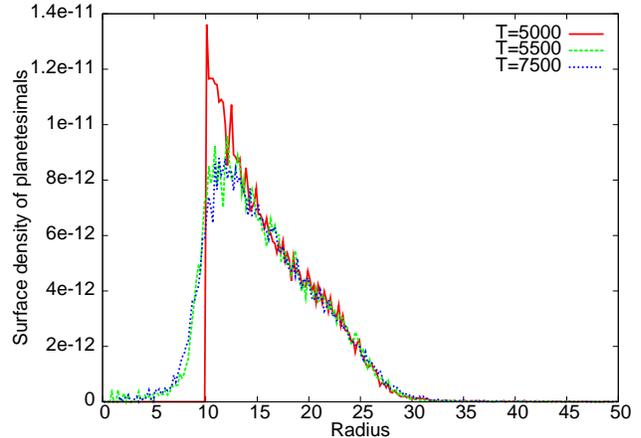}
\caption{Evolution of the planetesimals initially beyond R=10 at time t=5000.}
\label{Bulk}
\end{figure}

We do not need to conduct new simulations for this scenario but -
since the planetesimals are non-interacting - simply tag planetesimals
that are located at radius $R > 10$ at $t=5000$ and follow the evolution
of their density distribution. From Figure~\ref{Bulk}, one can see
that the distribution rapidly relaxes, with about 10 per cent of the
particles moving inward by a time of t=5500, although with no
discernible further evolution over the time-frame of $t=5500$ to
7500.  The initial relaxation is consistent with the fact that the
typical particle eccentricity is already $\sim 0.1$ at $t=5000$ and so
just represents the fact that particles near the `edge' visit the
inner region even without further orbital evolution. We revisit the
issue of planetesimal diffusion into an inner cleared region when we
construct our analytic model in Section 4.
  
\section{A simple scattering model}

The simulation results encourage the exploration of a scenario in
which the evolution of the planetesimal swarm is primarily driven by
changes in angular momentum rather than energy. This prompted us to
attempt to reproduce our results by modelling the evolution of the
swarm as a process of diffusion in angular momentum space. We start by
obtaining analytic expressions for the diffusion coefficients, set up
the resulting diffusion equation and use Green's functions to solve for
the evolution of the angular momentum distribution (and the associated
eccentricity and radial distributions) starting from an arbitrary
radial profile and initially circular orbits. The resulting solutions
however involve the computation of large numbers of Legendre
polynomials and are not of practical use until the system has evolved
well away from the initial delta function distribution of
eccentricities. We nevertheless obtain solutions for the eccentricity
distribution that are a reasonable representation of the SPH results
at the end of the simulation. We can also compute the subsequent
evolution of the system and derive expressions for the final
(equilibrium) distributions over radius and eccentricity (which
correspond to a uniform distribution in angular momentum at every
energy).  In the following sub-section we then verify these
predictions using a simple Monte Carlo model; this approach has the
additional advantage that it can be readily generalised to treat the
case where the perturbation amplitude is a function of radius.

\subsection{Analytic solution}
 We consider an ensemble of particles orbiting a mass M with fixed
 semi-major axis a. At some random time in each particle's orbit the
 velocity vector is rotated by some angle ${\rm \Delta \theta}$ in the
 orbital plane, where ${ \theta}$ is the angle between the velocity
 and radial vectors. This preserves the particle energy but perturbs
 the angular momentum.

We start by writing the angular momentum per unit mass, $L$, for a particle at
radius $r$ as 

\begin{equation}
 L^2=2GM\sin^2\theta(r-\frac{r^2}{2a}).
\label{eq:Lmax0}
\end{equation}

The equation is simplified by letting $ x=1-r/a$, so that

\begin{equation}
 L^2=GMa(1-x^2)\sin^2\theta=L_{{\rm max}}^{2}\sin^2\theta,
\label{eq:Lmax}
\end{equation}

and

\begin{equation}
 \frac{{\rm d}L}{{\rm d}\theta}=\sqrt{L_{{\rm max}}^{2}-L^2}.
\label{eq:DLL}
\end{equation}

$ L_{{\rm max}}$ is the maximum local angular momentum and is a
function of x (i.e. for particles at given $r/a$ and energy - hence
speed - this maximum occurs when the velocity is purely tangential).
The angular momentum diffusivity, $D$, is given by $\Delta L^2/\tau$
where $\Delta L$ is the change in angular momentum associated with
deflection $\Delta \theta$ and $\tau$ is the time between such
deflections: for now we assume both $\Delta \theta$ and $\tau$ to be
constant around the orbit.  We will also consider the case where
$\Delta \theta$ is independent of $a$ and where $\tau$ scales with the
orbital period ($\propto a^{3/2}$). This is motivated by the wish to
compare with the SPH simulations which are conducted with constant
cooling time to dynamical timescale ratio and in which, therefore, the
fractional amplitude of gravitational disturbances is independent of
radius (see equation (3)).
 
 We can relate $\Delta L$ to $\Delta \theta$ via Equation~\ref{eq:DLL}
 so that the value of $D$ at given $x$ is proportional to $({{\rm d}L}\over{{\rm d}\theta})^2$, that is to say 

\begin{equation}
 D \propto (\frac{{\rm d}L}{{\rm d}\theta})^2 \propto GMa(1-x^2)-L^2.
\label{eq:Lmax2}
\end{equation}

The expectation value of D over an entire orbit (at fixed $L$ and $a$)
is obtained by averaging this quantity over time, using the
relationship between time interval and r for an elliptical
orbit. Thus

\begin{equation}
\frac{{\rm d}r}{{\rm d}t}=\frac{\sqrt{GMa}}{r}\sqrt{e^2-(1-r/a)^2},
\label{eq:drdt}
\end{equation}

where $e$ is the orbital eccentricity, which is related
to $L$ and $a$ by

\begin{equation}
L^2=GMa(1-e^2).
\label{eq:L}
\end{equation}

Equation~\ref{eq:drdt} is derived from the orbital equation in the
appendix. As before, the equation is simplified by subsituting $x$ for
$r$, so that

\begin{equation}
\mathrm{{\rm d}}t \propto \frac{a^{3/2}(1-x)}{\sqrt{(e^{2}-x^2)}} \, \mathrm{d}x.
\label{eq:dxdt}
\end{equation}

Thus we have that (for fixed $a$)

\begin{equation}
\langle D \rangle  \propto \int_{-e}^{+e} \! \frac{(GMa(1-x^2)-L^2)(1-x)}{\sqrt{(e^2-x^2)}} \, \mathrm{d}x,
\label{eq:Dint}
\end{equation}

where the odd terms integrate to zero over symmetric limits and may be
discarded. The remaining even terms are standard integrals (requiring
the substitution ${x=e\sin w}$) and yield

\begin{equation}
\langle D \rangle \propto \frac{1}{2}(GMa-L^2).
\label{eq:DDDD}
\end{equation}

One can substitute this into Fick's Laws of diffusion to get the
following 1D diffusion equation for the angular momentum distribution
$h(L,t)$. $J$ is the flux and $A(a)$ is a constant that is determined
by the strength of the perturbation and which in general depends on
$a$ (see below).  In the case that we are currently considering (where
the fractional amplitude of perturbations in the disc is independent
of radius and where their characteristic timescale $\tau$ scales with
the orbital period i.e. $\propto a^{3/2}$) then A is proportional to
$a^{-3/2}$; under these circumstances the timescale for angular
momentum diffusion scales with the local orbital period.

 Thus we have

\begin{equation}
J=- \langle D \rangle \frac{\partial h}{\partial L},
\end{equation}

\begin{equation}
\frac{\partial h}{\partial t} = - \frac{\partial J}{\partial L},
\end{equation}

\begin{equation}
\langle D \rangle=A(a)(GMa-L^2).
\label{eq:aaa}
\end{equation}

The form of ${\langle D \rangle}$ contrasts with other
parametisations in the literature where it increases with
$L$. \citet{2009ApJ...701.1381A} consider, for example, the case where
it is proportional to $L$. The form of Equation~\ref{eq:aaa} reflects
the fact that where perturbations change only the direction of the
velocity, the associated angular momentum change goes to zero in the
limit of tangential orbits; this form thus prevents diffusion of
angular momentum beyond the physical limit set by a circular orbit.

The diffusion equation is then 

\begin{equation}
\frac{\partial h}{\partial t}=\frac{\partial}{\partial L}\biggl(A(a)(GMa-L^2)\frac{\partial h}{\partial L}\biggr),
\end{equation}

for which we assume a separable solution: ${h(L,t)=M(L)T(t)}$ to get

\begin{equation}
\frac{1}{T}\frac{{\rm d}T}{{\rm d}t} = \frac{1}{M}\frac{\rm d}{{\rm d}L}(A(a)(GMa-L^2)\frac{{\rm d}M}{{\rm d}L})=-k\, .
\end{equation}

This gives ${T(t) \propto {\rm exp}(-kt)}$. If $k$ is non-negative there will be no growing solutions. The equation for $M(L)$ is

\begin{equation}
\frac{\rm d}{{\rm d}L}((GMa-L^2)\frac{{\rm d}M}{dL}) + \frac{k}{A(a)}M =0 \,.
\label{eq:ML}
\end{equation}

Making the substitutions ${y=L/(\sqrt{GMa})}$ and ${k/A(a)=n(n+1)}$ converts Equation~\ref{eq:ML} into the Legendre equation:

\begin{equation}
\frac{\rm d}{{\rm d} y}((1-y^2)\frac{{\rm d}M}{{\rm d}y}) + n(n+1)M =0.
\end{equation}

The solutions must be regular at ${\rm y=\pm1}$, so we may use the Legendre polynomials. If this is combined with the solution for $t$ we have

\begin{equation}
h_n(L,t)=c_n{\rm exp}(-n(n+1)A(a)t)P_n(L/\sqrt{GMa}).
\end{equation}

The general solution is a linear combination of the ${h_n}$ solutions, with the coefficients determined by the initial distribution: 

\begin{equation}
h(L,t)=\sum_{n=0}^{\infty} c_n{\rm exp}(-n(n+1)A(a)t)P_n(L/\sqrt{GMa}).
\end{equation}

If the initial distribution consists of $N$ particles in prograde
circular orbits, we have ${h(L,t=0)=N\delta(L-\sqrt{GMa})}$. If we use
the orthogonality property of the Legendre polynomials and the fact
that ${P_n(1)=1}$ for all n, we have

\begin{equation}
h(L,t)=\frac{N}{\sqrt{GMa}}\sum_{n=0}^{\infty} \frac{2n+1}{2} {\rm exp}(-n(n+1)At)P_n(L/\sqrt{GMa}).
\end{equation}

The eccentricity distribution is then obtained by using $f(e){\rm d}e=h(L){\rm d}L+h(-L){\rm d}L$. This is because an eccentricity of $e$ may be owing to a prograde or retrograde orbit with angular momenta of the same magnitude. The quantity d$L$ is double-valued for the same reason. Thus from Equation~\ref{eq:L} we have: 

\begin{equation}
L=\pm \sqrt{GMa(1-e^2)},
\end{equation}

and

\begin{equation}
\mathrm{d}L=\frac{\mp e\sqrt{GMa}}{\sqrt{1-e^2}} \, \mathrm{d}e.
\end{equation}

The Green's function solution, $f(e,t,a)$ for
initially circular orbits at a particular semi-major axis $a$ is therefore

\begin{equation}
f(e,t,a)=\frac{Ne}{\sqrt{1-e^2}}\sum_{n=0}^{\infty} (2n+1) {\rm exp}(-n(n+1)A(a)t)P_n(\sqrt{1-e^2}).
\end{equation}

\begin{figure}
\centering
\includegraphics[width=0.99\columnwidth]{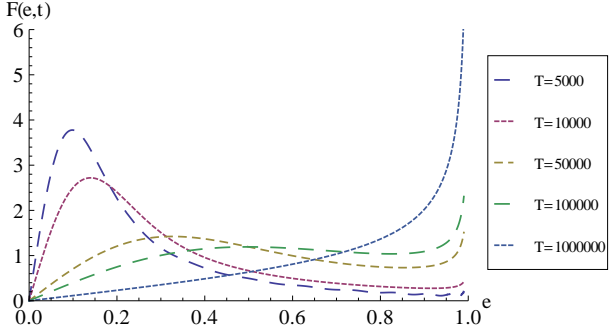}
\caption{Evolution of the normalised eccentricity distribution.}
\label{F(e)}
\end{figure}

To consider the behaviour of an entire disc of particles we then
calculate an appropriate superposition of Green's functions, bearing
in mind (as argued above) that $A=A_o a^{-3/2}$.  For comparison with
the SPH simulations, we take a number distribution per unit $a$ that
is proportional to ${a^{-1/2}}$ (since this corresponds to the
initial surface density distribution scaling as $R^{-3/2}$) and adopt
inner and outer radii of $a_1=1$ and $a_2=25$. The integrations are
straightforward but laborious, making Mathematica the obvious
choice. We plot $F(e,t)$ (the distribution function for particle
eccentricity) in Figure~\ref{F(e)} and have normalised the time unit
so as to match the peak in the distribution obtained from the SPH
simulation at a time $t$=5000. At $t$=0, $F(e,t)$ is a delta function at $e$=0
but this and the subsequent evolution before $t$=5000 required too many
Legendre polynomials to be computationally practicable. We however
note the similar form of the curve at $t$=5000 to that obtained in the
SPH simulation at the same time (Figure 7). The final distribution
occurs when $t$ is large enough such that all the non-constant Legendre
terms are effectively zero, leaving ${F(e,t) \propto
  e/\sqrt{1-e^2}}$. This is an equilibrium state and represents a
uniform angular momentum distribution for particles with the same
semi-major axis.

The surface density distribution can be obtained through similar
means. Equation~\ref{eq:dxdt} gives the probability of a particle with
eccentricity $e$ being at $x$. Multiplying this by $f(e,t,a)$ and
integrating with respect to $e$ gives $n(x,t,a$), the fraction of
particles at given semi-major axis in a given interval of $x$.The
lower limit is the value of $e$ for an orbit that achieves the required
$x$ value at either periapsis or apoapsis.

\begin{equation}
n(x,t,a) \propto \int_{max(x,-x)}^{1} \! f(e,t)\frac{1-x}{\sqrt{e^{2}-x^{2}}} \, \mathrm{d}e.
\end{equation}

This must be converted to a function of $r$ and $a$ and then normalised so
that it represents the distribution resulting from a fixed number of
particles. It can then be integrated with the power law distribution
for $a$ to get the overall number density.

\begin{equation}
N(r,t) \propto \int_{a_{1}}^{a_2} n(r,t,a)a^{-1/2} \, \mathrm{d}a.
\label{eq:numberd}
\end{equation}
\begin{equation}
\Sigma(r,t)= \frac{N(r,t)}{2\pi r}.
\label{eq:sigmad}
\end{equation}

Again, the integrations are complicated for very many Legendre
polynomials.  However the solution for the equilibrium distribution is
readily obtained.  We note that (for a particle swarm of fixed $a$),
the time spent in the interval $r$ to $r+{\rm d}r$ is
$({{{\rm d}r}\over{{\rm d}t}})^{-1}{\rm d}r$, which (using Equation~\ref{eq:drdt} and
Equation~\ref{eq:Lmax0}) can be written (for particles of angular
momentum $L$) as

\begin{equation}
{\rm d}t \propto {{r {\rm d}r}\over{\sqrt{L_{{\rm max}}^2-L^2}}}.
\end{equation}

 Integrating over all values of $L$ from $-L_{{\rm max}}$ to $L_{{\rm max}}$ 
we obtain (in general):

\begin{equation}
{\rm d}t \propto r {\rm d}r  \int_{-L_{{\rm max}}}^{L_{{\rm max}}}{{h(L){\rm d}L}\over{\sqrt{L_{{\rm max}}^2-L^2}}}.
\end{equation}

In equilibrium, h is constant which means that one may write the integral
in terms of the dimensionless variable $\tilde L = L/L_{max}$ so that

\begin{equation}
{\rm d}t \propto r {\rm d}r  \int_{-1}^{+1} {{{\rm d}\tilde L}\over{\sqrt{1-\tilde L^2}}}.
\end{equation}

 We thus find that in equilibrium the fraction of particles between
$r$ and $r+{\rm d}r$ scales as $r {\rm d}r$. Now at fixed $a$, the minimum and maximum
radii attained by the particle are $0$ and $2a$ so that the normalisation
is such that 

\begin{equation}
n_{eq}(r,a)= {{rdr}\over{2a^2}} \, (r < 2a).
\end{equation}

  We can now find the surface density density distribution by substituting
$n(r,t,a) = n_eq(r,a)$ into Equations~\ref{eq:numberd} and ~\ref{eq:sigmad}. Thus

\begin{equation}
\Sigma(r,t) \propto \frac{N(r,t)}{2\pi r} \propto \frac{1}{r}\int_{a_{1}}^{a_2} r a^{-5/2} \, \mathrm{d}a \, (a>r/2).
\end{equation}

 This gives rise to three regimes. For $ 0 < r < 2a_1$, all values of
$a$ contribute to the integral throughout this range and the surface
density is {\it constant}. For $2a_1 < r < 2a_2$ then it is only $a$
values greater than $r/2$ which contribute; consequently the surface
density falls with radius as the annulus at $r$ is populated by
an ever decreasing range of $a$ values. Finally, for
$r > 2 a_2$, no particles can be scattered to this radius and the
surface density is zero.

Summarising we therefore have:
 
\begin{equation}
\Sigma(r)=A \, (0<r<2a_1),
\end{equation}
\begin{equation}
\Sigma(r) =  A {{(1 - (r/a_2)^{-3/2}}\over{1 - (a_1/a_2)^{-3/2}}} \,  (2a_1<r<2a_2),
\end{equation}
\begin{equation}
\Sigma(r)=0 \, (r>2r_2).
\end{equation}

  We note that although we have derived the above for a specific
distribution of particles in $a$, it is a {\it general}
result that - in equilibrium - the surface density is
uniform for radii less than twice the inner radius of the initial
distribution. The reason for this is that every group of particles
of given $a$ is redistributed, in equilibrium, into
a uniform density disc of particles extending from zero to $2a$. At
every radial position inward of $2 a_1$, {\it all} bins of $a$ 
provide uniform surface density contributions to the total surface
density. It is only at radii $> 2 a_1$ that particles with
$a < r/2$ cease to contribute and the surface density starts to decline.

  We emphasise that this entire derivation is based on the assumption
that the disc fluctuations are scale-free, which implies that
$\Delta \theta$ is independent of $a$ and independent of phase at 
given $a$. In reality, however, the amplitude of perturbations
in self-gravitating discs is an increasing function of radius. This
changes the analysis in two ways: it changes the dependence of
$D$ on $L$ (cf Equation~\ref{eq:DDDD}) because the perturbations are of
larger amplitude near apocentre and it also changes the scaling
of $D$ with $a$. As an example, if ${\rm ({\rm d}\theta)^2 \propto r^p}$
then the diffusion coefficient becomes (by analogy with Equation~\ref{eq:Dint}):

\begin{equation}
\langle D \rangle  \propto a^p\int_{-e}^{+e} \! \frac{(GMa(1-x^2)-L^2)(1-x)^{p+1}}{\sqrt{(e^2-x^2)}} \, \mathrm{d}x.
\end{equation}

 In general, this integral is not analytic, except where p=-1: in this
 case the removal of the term ${ (1-x)}$ makes no difference since
 the term $\propto x$ vanishes by symmetry for $p=0$. Consequently the
 evolution at fixed $a$ still reduces to the Legendre equation,
 although the computation for a range of $a$ now needs to take account
 of the additional $a$ dependence of $D$. We do not pursue this
 further here, since - in considering situations that are likely to
 occur in real discs - we want to be able to treat cases other than
 $p=-1$. This forms part of our motivation for adopting a Monte Carlo
 method in the following section.

\subsection{Monte Carlo simulations}

\begin{figure}
\centering
\includegraphics[width=0.99\columnwidth]{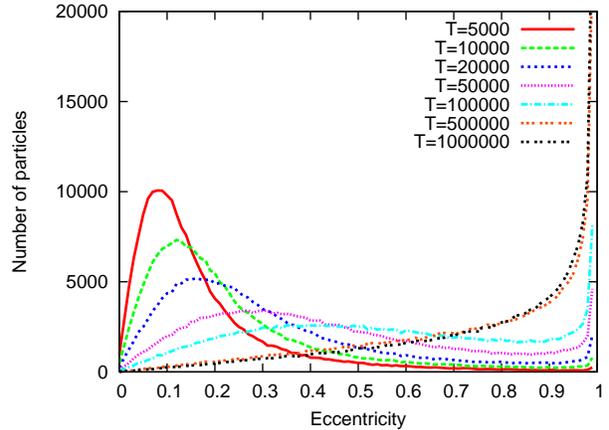}
\caption{Evolution of the un-normalised eccentricity distribution from the Monte Carlo simulation.}
\label{FinalE}
\end{figure}

  We have checked and extended the above analysis of the behaviour of
  particle swarms that undergo diffusion in angular momentum at fixed
  energy by conducting simple Monte Carlo simulations. Here we subject
  co-planar particles (initially in circular orbits with the same
  surface density profile as employed in the SPH simulations) to
  random rotation of their velocity vectors through an angle in the
  range $\pm\Delta \theta$ at a random time each orbit.  As expected,
  the evolution is qualitatively independent of the value of $\Delta
  \theta$ and with a timescale that scales as $\Delta \theta^{-2}$. We
  have adjusted our parameters so as to match the peak of the
  eccentricity distribution in the SPH calculation at a time $t=5000$
  and verified that the evolution of the eccentricity distribution
  (Figure~\ref{FinalE}) is in good agreement with the output of our
  analysis of the diffusion equation (Figure~\ref{F(e)}). Secondly,
  the early evolution compares reasonably well with the f(e) curves
  obtained from the SPH simulation (Figure~\ref{E}). However, the
  equilibrium distribution for f(e) is only reached after several
  hundred thousand inner orbital periods.

\begin{figure}
\centering
\includegraphics[width=0.99\columnwidth]{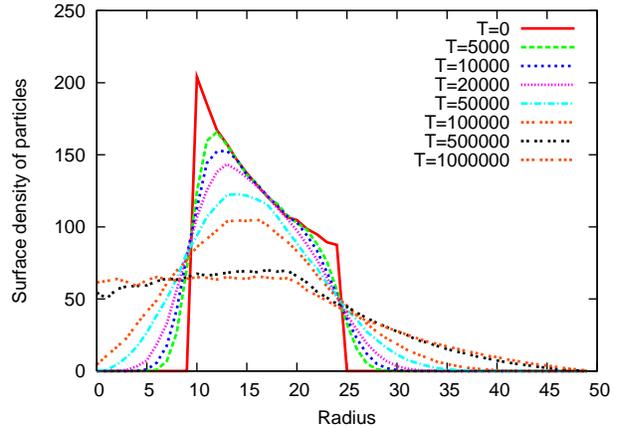}
\caption{Evolution of the surface density distribution, in arbitrary units, from the Monte Carlo simulation.}
\label{FinalSD}
\end{figure}

We also looked at the evolution of the surface density distribution of
the planetesimals with initial radii greater that R=10
(Figure~\ref{FinalSD}), so as to compare with that for the SPH
simulation (Figure~\ref{Bulk}). In the Monte Carlo simulation, the
surface density profile evolves through a state similar to that seen
in the SPH results and eventually attains a true equilibrium, where
the surface density is approximately flat out to twice the inner
radius (R=20)and declines to zero at twice the outer radius
(R=50). This is the result expected from the earlier analysis.

\subsection{Application to real discs}

 We have hitherto adopted the assumption that the fractional amplitude
 of perturbations - and hence the value of $\Delta \theta$ - is
 independent of radius. We have also normalised the evolution of the
 planetesimal swarm to that found in the SPH simulations which adopted
 rather short cooling times ($\beta = 5-10$) and correspondingly large
 amplitude of gravitational instability (with fractional amplitudes of
 several tens of percent \citep{2010MNRAS.401.2587C}). This choice was
 motivated by the need to compute the evolution of the planetesimal
 swarm in a reaonable time and also to ensure a sufficiently vigorous
 gravitational instability for it not to be quenched by numerical
 viscosity \citep{2011MNRAS.413.2735L}.  The insights provided by the
 simulations can now be used to rescale the problem to the parameter
 range expected in real self-gravitating discs.

 We therefore consider the case of an optically thick self-gravitating
 disc accreting at $3 \times 10^{-6} M_\odot$ yr$^{-1}$ for which the
 steady state solutions (employing opacities due to dust and gas from
 \citet{1994ApJ...427..987B}) are detailed in the Appendix of
 \citet{2009MNRAS.396.1066C}.  At radii $> 35$ au, the opacity is
 dominated by ice and we have:

\begin{equation}
\beta = 3 \times 10^4 \biggl({{R}\over{10 \rm au}}\biggr)^{-9/2}.
\end{equation}

 For $22 {\rm au} < R < 35 {\rm au}$ (where opacity is dominated by
ice sublimation) and

\begin{equation}
\beta = 270 \biggl({{R}\over{10 \rm au}}\biggr)^{-9/14},
\end{equation}

while for $4 {\rm au} < R < 22 {\rm au}$

\begin{equation}
\beta = 1000 \biggl({{R}\over{10 \rm au}}\biggr)^{-9/4}.
\end{equation}

 Since the amplitude of pertubations scales as $\beta^{-1/2}$ we also
 expect that $\Delta \theta \propto \beta^{-1/2}$; we normalise the
 value of $\Delta \theta$ at $\beta = 5$ in order to match the rate of
 evolution of the SPH simulation.  Within $R=4$ au we set $\Delta
 \theta = 0$ since it is arguable whether the disc is self-gravitating
 at this point; in any case, the long cooling times there equate to
 negligibly small values of $\Delta \theta$.
 
 We start with a belt of planetesimals in circular orbits which are
 distributed with a power law surface density distribution with
 $\Sigma \propto R^{-3}$ since this is the steady state gas surface
 density profile for a self-gravitating disc with opacity dominated by
 ice grains \citep{2009MNRAS.396.1066C}. The belt extends from $60$ to
 $100 $ au, the choice of inner radius being motivated by the fact
 that rather large amplitude density enhancements (rapid cooling) are
 required in order to {\it form} planetesimals through dust
 concentration in spiral arms \citep{2009MNRAS.398L...6C}.  A primary
 motivation for modeling a distribution with an initial hole is to
 discover whether - with a realistic prescription for the radial
 dependence of the perturbation amplitude - one expects planetesimals
 to be scattered inwards to fill up the hole over the self-gravitating
 lifetime of the disc.

\begin{figure}
\centering
\includegraphics[width=0.99\columnwidth]{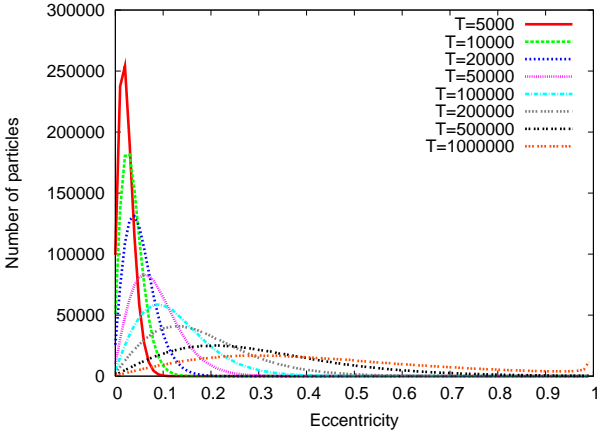}
\caption{Evolution of the un-normalised eccentricity distribution from the Monte Carlo simulation with realistc cooling.}
\label{RealE}
\end{figure}

\begin{figure}
\centering
\includegraphics[width=0.99\columnwidth]{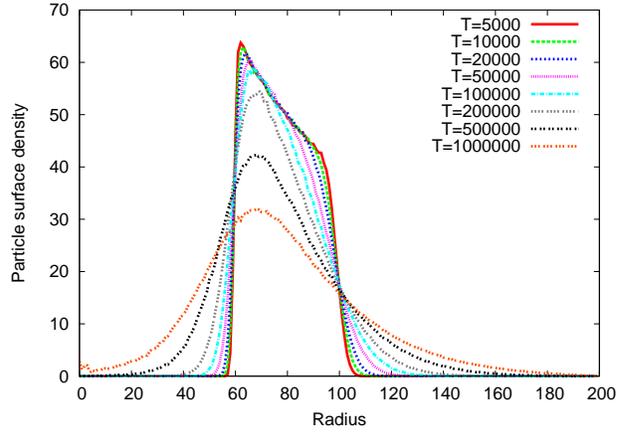}
\caption{Evolution of the surface density distribution, in arbitrary units, from the Monte Carlo simulation with realistc cooling.}
\label{RealSD}
\end{figure}

  We find (Figure~\ref{RealSD}) that the evolution is qualitatively similar to that
  in Figure~\ref{FinalSD}: note that the edge is at 60 au in this simulation, as
  opposed to 10 code units in Figure~\ref{FinalSD}. In both cases, the time is
  normalised to the dynamical time at the inner edge (1 code unit and
  1 au respectively); therefore in order to compare the two plots at
  a given number of orbital times {\it at the truncation point} it is
  necessary to multiply the times in Figure~\ref{FinalSD} by $6^{1.5}$.  It is
  then evident that the timescale for the infilling of the hole is
  rather similar in the two cases. i.e. the rate of infill is
  controlled by the amplitude of fluctuations {\it near the truncation
    radius}. This can be understood inasmuch that particles that visit
  the inner disc on eccentric orbits spend most of their time near
  apocentre (i.e.  close to the truncation radius). The fact that - in
  the realistic variable $\beta$ case - the spiral structure is of
  very low amplitude in the inner disc therefore has little effect on
  the evolution of the planetesimal swarm.

\section{Conclusions}

 The SPH simulations indicate that planetesimal eccentricity is
 efficiently amplified by interaction with spiral features in
 self-gravitating discs. In the case of discs where the cooling time
 is not much longer than the dynamical timescale (i.e in the outer
 disc, at radii $> 10$s of au, where we expect planetesimals to be
 formed in such discs), the amplitude of these features is sufficient
 for high eccentricities ($> 0.1$) to be driven on much less than the
 self-gravitating lifetime of the disc. We have found that there is no
 significant velocity coherence within the particle swarm and thus
 that the local velocity dispersion is a significant fraction of the
 local orbital velocity; we also find that there is no tendency - in
 the test particle regime studied here - for the planetesimals to be
 concentrated within spiral arms. The lack of significant density
 enhancements and the large velocity dispersion (which weakens
 gravitational focusing) both contribute to very long collision times
 ($>$ a Gyr); moreover, any collisions that did occur would be in the
 destructive regime.

 We have used the SPH simulations to calibrate Monte Carlo experiments
in which the particle direction is randomly perturbed around its orbit
and have compared these Monte Carlo experiments with an analytical
description.  The Monte Carlo experiments allow the long time
integration of the system and can also probe the high cooling time
(weak spiral) regime that cannot be reliably simulated hydrodynamically.
We have treated the case of an initial planetesimal belt at large
radius ($60$ au) where the radial variation of the spiral
amplitude is parametrised in terms of expected variations in the
local cooling physics in marginally unstable self-gravitating discs.
We find that - notwithstanding the fact that the spiral potential
is very weak at small radius - there is significant scattering of
planetesimals into the inner disc, with particles that are
perturbed in the region beyond $60$ au visiting small radii on
eccentric orbits. We nevertheless find that most of the particles
initially beyond $60$ au are likely to be retained at large radius
on timescales of $\sim 10^5$ years.

 The picture that emerges from this study is that if planetesimals
do form in the self-gravitating phase of disc evolution then they
will be retained in the disc throughout this phase;  their
high eccentricities inhibit collisions and mean that they will neither
grow nor suffer significant collisonal disruption over this period.
If (as argued by \citet{2009MNRAS.398L...6C}), such planetesimals are
only formed in the outer disc then they will be largely retained at
such radii, though with a significant minority that are perturbed to the
inner regions of the disc. Such an endpoint would therefore represent
the initial conditions for considering the subsequent evolution of
planetesimals during the non-self gravitating phase of disc evolution.
  
\section*{Acknowledgements}
JJW thanks the STFC for his studentship. We would like to thank
Guiseppe Lodato for providing much valuable advice, Mark Wyatt and Jim
Pringle for discussion and comments, and John Eldridge for
proof-reading the paper.

\bibliographystyle{mn2e}
\bibliography{planet}

\appendix

\section{Integrating around an orbit}

The expected value of D is obtained by averaging over one orbit.  This requires obtaining dt as a function of r. We start with the orbital equation for r as a function of $\phi$.

\begin{equation}
\rm r=\frac{a(1-e^2)}{1+e\cos\phi}
\label{eq:polar}
\end{equation}

\begin{equation}
\rm \frac{dr}{dt}=\frac{dr}{d\phi}\frac{d\phi}{dt}=\frac{dr}{d\phi}\frac{L}{r^2}
\end{equation}

L is a constant of the motion for given values of a and e.

\begin{equation}
\rm L^2=GMa(1-e^2)
\label{eq:Lappen}
\end{equation}

\begin{equation}
\rm \frac{dr}{dt}=\frac{a(1-e^2)e\sin\phi}{(1+e\cos\phi)^2}\frac{\sqrt{GMa(1-e^2)}}{r^2}
\end{equation}

Substituting r back in simplifies the expression.

\begin{equation}
\rm \frac{dr}{dt}=\frac{\sqrt{GMa}}{1-e^2}\frac{\sin\phi}{a}
\end{equation}

Re-arranging Equation~\ref{eq:polar} and using that to eliminate $\phi$ gives an expression in r and e only.

\begin{equation}
\rm \frac{dr}{dt}=\frac{\sqrt{GMa}}{r}\sqrt{e^2-(1-r/a)^2}
\end{equation}

\end{document}